\def\rfr#1{eq. (\ref{#1})}

\def\bb#1#2#3{\bibitem[\protect\citeauthoryear{#1}{#2}]{#3}}
\def\eqi{\begin{equation}}
\def\eqf{\end{equation}}
\def\eqia{\begin{eqnarray}}
\def\eqfa{\end{eqnarray}}
\def\rp#1#2{{#1\over#2}} \def\lb#1{\label{#1}}
\def\bb#1#2#3{\bibitem[\protect\citeauthoryear{#1}{#2}]{#3}}

\def\bds#1{\boldsymbol{#1}}

\documentclass{aastex}
\usepackage{url}
\usepackage{amsmath,amsthm,amscd,amssymb}
\usepackage{latexsym,wasysym}
\usepackage{graphicx,epsfig}

\RequirePackage{color}

\newcommand{\emaila}{lorenzo.iorio@libero.it}

\begin{document}

\title{Impact of a Pioneer/Rindler-type acceleration on the Oort cloud}

\author{Lorenzo Iorio\altaffilmark{1} }
\affil{Ministero dell'Istruzione, dell'Universit\`{a} e della Ricerca (M.I.U.R.)-Istruzione\\
Fellow of the Royal Astronomical Society (F.R.A.S.)\\
International Institute for Theoretical Physics and
High Mathematics Einstein-Galilei\\
 Permanent address for correspondence: Viale Unit\`{a} di Italia 68, 70125, Bari (BA), Italy}

\email{\emaila}

\begin{abstract}
According to a recent modified model of gravity at large distances,  a radial constant and uniform extra-acceleration $\bds A_{\rm Rin }=A_{\rm Rin}\bds{\hat{r}}$ of Rindler type acts upon a test particle p in the static field of a central mass $M$ if certain conditions are satisfied. Among other things, it was proposed as a potentially viable explanation of a part of the Pioneer anomaly.
We study the impact that  an anomalous Rindler-type  term as large as $|A_{\rm Rin}|\sim 10^{-10}$ m s$^{-2}$ may have on the the orbital dynamics of a typical object of the Oort cloud whose self-energy is quite smaller than its putative Rindler energy. By taking a typical comet moving along a highly eccentric and inclined orbit throughout the expected entire extension of the Oort cloud ($\sim 0.02\ {\rm pc}-1$ pc), it turns out that the addition of an outward Rindler-like acceleration, i.e. for $A_{\rm Rin}>0$, does not allow bound orbits. Instead, if $A_{\rm Rin}<0$, the resulting numerically integrated trajectory is limited in space, but it radically differs from the standard Keplerian ellipse. In particular, the heliocentric distance of the comet gets markedly reduced and experiences high frequency oscillations, its speed is increased, and the overall pattern of the trajectory is quite isotropic. As a consequence, the standard picture of the Oort cloud is radically altered since its modified orbits are much less sensitive to the disturbing actions of the Galactic tide and nearby passing stars whose effects, in the standard scenario, are responsible for the phenomenology on which our confidence in the existence of the cloud itself is based. The present analysis may be supplemented in future by further statistical Monte Carlo-type investigations by randomly varying the initial conditions of the comets.
\end{abstract}


\section{Introduction}\lb{intro}
Recently, \citet{Gru010}, under certain assumptions, put forth a quite general model for the gravitational field of a static central object of mass $M$ at large distances $r$ from it. As a result, the acceleration felt by a test particle p in the field of $M$ turns out to be modified by the appearance of a Rindler-type additional term. Thus, the total acceleration becomes  \citep{Gru010,Carlo011,GruPre011}
\eqi\bds{A}=-\rp{GM}{r^2}\bds{\hat{r}}+\bds A_{\rm Rin},\lb{tutta}\eqf
where $G$ is the Newtonian gravitational constant, and
\eqi \bds A_{\rm Rin}= A_{\rm Rin}\bds{\hat{r}}:\lb{grum}\eqf
the sign of $A_{\rm Rin}$  is left indeterminate by the theory.
Notice that \citet{GruPre011} and \citet{Culetu011} assumed that $A_{\rm Rin}$ is a universal constant.
Concerning its sign and magnitude, \citet{GruPre011} and \citet{Culetu011} took
\eqi \left|A_{\rm Rin}\right|= 1\times 10^{-10}\ {\rm m\ s^{-2}}\lb{magni};\eqf \citet{GruPre011} assumed an inward direction.

Importantly, \citet{Carlo011,GruPre011} argued that, if \rfr{tutta} has to be valid for a given physical system, then
the following condition
\eqi \rp{Gm_{\rm p}}{d_{\rm p}}\lesssim \left|A_{\rm Rin}\right|r\lb{condizione}\eqf
must be fulfilled. In \rfr{condizione} $m_{\rm p}$ and $d_{\rm p}$ are the mass and a typical size of the test particle p, respectively. Otherwise,  the self-energy of the test particle would overwhelm the Rindler energy, and the consequent particle's backreaction on the background would not be negligible. In this case, it would not be possible to assume the universal (=maximal) value for the Rindler acceleration. For more details, see the discussion in Section V of \citet{Carlo011}. Actually, it is easy to show that \rfr{condizione} is not valid for the planets of the solar system
since it is
\eqi \rp{Gm_{\rm p}}{d_{\rm p} \left|A_{\rm Rin}\right| r}= \rp{(6\times 10^2-1\times 10^5)\ {\rm kau}}{r},\eqf
where 1 kau = 1000 astronomical units = $4.85\times 10^{-3}$ pc. \citet{Carlo011,GruPre011} noticed that \rfr{condizione} is, instead, fully satisfied for the Pioneer 10/11 probes, affected by the well known Pioneer anomaly \citep{And98,And02} which may partly be  explained by the Rindler-type acceleration \citep{Gru010,Carlo011,GruPre011}. \citet{Carlo011} obtained the following constraint
\eqi |A_{\rm Rin}|\lesssim 3\times 10^{-9}\ {\rm m\ s^{-2}}\eqf
from the propagation of electromagnetic waves, for which the previous caveat is not a concern.

Actually, the solar system host several natural objects orbiting the Sun for which \rfr{condizione} holds: they  reside in the Oort cloud \citep{Opik32,Oort50}. It is a reservoir of frozen cometary nuclei which is supposedly located in the remote peripheries of the solar system. It should be a remnant of the early stages of the formation of the solar system, and it likely formed as a consequence of scattering of planetesimals by the giant planets \citep{Dun87,Higu06}. The Oort cloud  has  likely a  spheroidal shape \citep{Weiss96} and a size ranging from about $5-10$ kau up to  $150-200$ kau \citep{Levi07}. Its existence was conjectured by noticing that the typical lifetimes of comets near the Sun is of the order of $10^4-10^5$ yr \citep{Levi94} due to non-gravitational (sublimation) and gravitational (strong  interactions with planets) phenomena, while the age of the solar system is $4-5$ orders of magnitude larger. Thus, a continuous resupplying from a much more remote, longer-lived source should take place in order to maintain the cometary population in a steady state, as it is observed. Long-period\footnote{The
threshold of 200 yr was chosen mainly for historical reasons: it is arbitrary \citep{Morbi05}.} ($P_{\rm b}\geq 200$ yr)  comets, characterized by highly inclined and eccentric orbits \citep{Morbi05}, would originate just from the Oort cloud \citep{Morbi05,Dun08}. They are injected into observable orbits in the planetary regions of the solar system by gravitational interactions with interstellar medium \citep{Stern90} and giant molecular clouds \citep{Maze04,Jacu08}, nearby passing stars \citep{Hills81,Boby010a,Boby010b} and the Galactic tide (radial and vertical) \citep{Rick08,Masi09}: as a result, comet showers may occur \citep{Heis87,Mate95}. Also other existing minor objects of the solar system like Centaurs, highly elliptical trans-Neptunian objects and Jupiter-family comet population may come from the Oort cloud \citep{Emel07}.
 Connections of terrestrial cratering with such cometary showers originating from the Oort cloud have been investigated by some researchers \citep{Wick08}.

By taking the Halley\footnote{Although $P_{\rm b}=75.3$ yr, its orbital characteristics point towards a capture from the long-period population of the Oort cloud \citep{Ferna02}.} comet as representative, a typical Oort object p  may be thought as characterized by a mass \citep{Cevo87}
\eqi m_{\rm p}=2.2\times 10^{14}\ {\rm kg}\eqf
and   mean size \citep{Lamy04}
\eqi d_{\rm p}=11\ {\rm km},\eqf
with a mean density of \citep{Sag88}
\eqi\rho_{\rm p} = 0.6\ {\rm g\ cm^{-3}},\eqf
although uncertainties in several parameters and assumptions \citep{Peale89} may lead to a range of values as large as
\eqi \rho_{\rm p}=0.2-1.5\ {\rm g\ cm^{-3}}.\eqf
Such figures and \rfr{magni} imply that
\eqi \rp{Gm_{\rm p}}{d_{\rm p} \left|A_{\rm Rin}\right| r}=\rp{0.0892\ {\rm au}}{r}.\eqf
Thus, the condition \rfr{condizione} is fully satisfied over the entire extension of the Oort cloud also for bodies which can be much larger and denser than a typical cometary nucleus. Indeed, \rfr{condizione} can approximately be posed as
\eqi \rp{\pi G\rho_{\rm p} d^2_{\rm p}}{\left|A_{\rm Rin}\right| r}\lesssim 1;\eqf
for, say, $\rho_{\rm p} = 5$ g cm$^{-3}$, which is the mean density\footnote{Clearly, it is highly unlikely that  such dense bodies can really exist in the Oort cloud.} of the rocky planets \citep{McFad07}, and $d_{\rm p}=250$ km we would have
\eqi \rp{\pi G\rho_{\rm p} d^2_{\rm p}}{\left|A_{\rm Rin}\right| }=4.4\ {\rm kau}.\eqf

In this paper we will explore the consequences that the existence of a Rindler-type extra-acceleration, with the characteristic of \rfr{grum} and \rfr{magni}, would have on the orbital motion of a typical Oort cloud object. Should the resulting orbital pattern  radically differ from the standard Newtonian one, shadows on the Rindler-like acceleration would be casted since the entire dynamical history of the Oort cloud should be re-written and all the inferences nowadays accepted, based on the Newtonian picture of the Oort cloud and of its interaction with the surrounding stellar and Galactic environment, would not be valid anymore. Actually, non-negligible modifications of the Newtonian orbits are expected since the mean Newtonian accelerations $A_{\rm N}$ of an Oort comet may range from $A_{\rm N}=2\times 10^{-10}\ {\rm m\ s^{-2}}\ (r=5\ {\rm kau})$ to $A_{\rm N}=1.5\times 10^{-13}\ {\rm m\ s^{-2}}\ (r=200\ {\rm kau})$.
Concerning general relativity, it will not be considered in the rest of the paper since its effects are totally negligible in the present context. Suffices it to say that the 1PN term of order $\mathcal{O}(c^{-2})$ causes a comet acceleration as small as \eqi A_{\rm 1PN}\sim \rp{(GM)^2}{c^2 r^3}\lesssim 10^{-21}\ {\rm m\ s}^{-2},\eqf where $c$ is the speed of light in vacuum.
As we will see in Figure \ref{figura2} (Section \ref{sec.2.1}), no special relativistic effects will come into play since, although generally increased by the Rindler acceleration, the speed of the Oort comet will remain several orders of magnitude smaller than $c$.

The plan of the paper is as follows. In Section \ref{sec.2} we numerically compute the trajectory of a typical Oort object for inward (Section \ref{sec.2.1}) and outward (Section \ref{sec.2.2}) directions of $\bds A_{\rm Rin}$. In Section \ref{sec.3} we investigate if an Oort comet acted upon by a Rindler-type acceleration is sensitive to the perturbing actions of the Galactic tide (Section \ref{sec.3.1}) and of nearby passing stars (Section \ref{sec.3.2}) which, in the standard Newtonian scenario, cause the phenomenology on which our confidence in the existence of the Oort cloud is based. In Section \ref{sec.4} we summarize and discuss our findings.
\section{Numerically produced orbits of  Oort comets affected by the Rindler acceleration }\lb{sec.2}
\subsection{An inward Rindler acceleration}\lb{sec.2.1}
The orbital effects of \rfr{grum}-\rfr{magni} on an Oort comet p cannot be computed perturbatively since, according to \rfr{magni}, the magnitude of $A_{\rm Rin}$ is of the same order of magnitude of, or larger than the Newtonian acceleration experienced by p.

Thus, we numerically integrate the modified equations of motion of p in a heliocentric  frame endowed with cartesian coordinates. We adopt initial conditions, shown in Table \ref{tavola}, such that, in the limit $A_{\rm Rin}\rightarrow 0$, the resulting orbit reduces to a standard Keplerian ellipse. It is characterized by large values of its  eccentricity $e$ and inclination $I$ to the reference $\{x,y\}$ plane, and the semi-major axis $a$ is chosen in such a way that the comet's motion covers almost all the expected extension of the Oort cloud. The other standard Keplerian orbital elements are the longitude of the ascending node $\mathit{\Omega}$, the argument of pericenter $\mathit{\omega}$ and the true anomaly $f$. {In order to facilitate a comparison with other studies on the Oort cloud,  we adopted as reference frame a heliocentric inertial one having the  $x$ axis directed towards the Galactic Center, the $y$ axis directed along the Sun's Galactic velocity, and the $z$ axis completing
a right-handed system\footnote{The Galactic plane is tilted by  $62.6$ deg to the celestial equator \citep{Sulli84}, so that it is almost at right angle to the ecliptic.} \citep{Fouch05}.}
\begin{table*}[ht!]
\caption{Initial conditions, in kau and ${\rm kau\ Myr^{-1}=4.74\ m\ s^{-1}}$, adopted for the numerical integration of the modified equations of motion of an Oort comet acted upon by the Rindler-type acceleration of \rfr{grum}. In terms of standard Newtonian mechanics, they correspond to a Keplerian ellipse with $a=102.507$ kau, $e=0.955$, $I=77.41$ deg, $\mathit{\Omega}=-139.87$ deg, $\mathit{\omega}=321.94$ deg, $f=-152.14$ deg. {They can be thought as referred to the heliocentric fixed frame of Figure 1 in \citet{Fouch05}.} The orbital period $P_{\rm b}= 32.81$ Myr. The perihelion distance is $q\doteq a(1-e)=4.606$ kau, and the aphelion distance is $Q\doteq a(1+e)=200.407$ kau.
}\label{tavola}
\centering
\bigskip
\begin{tabular}{llllll}
\hline\noalign{\smallskip}
$x_0$ (kau) & $y_0$ (kau) & $z_0$ (kau) &  $\dot x_0$ (kau Myr$^{-1}$) & $\dot y_0$ (kau Myr$^{-1}$) & $\dot z_0$ (kau Myr$^{-1}$) \\
\noalign{\smallskip}\hline\noalign{\smallskip}
$45$ & $35$ & 10 & $-23$ &  $-15$ & $-15$\\
\noalign{\smallskip}\hline\noalign{\smallskip}
\end{tabular}
\end{table*}
The results of the integrations are displayed in Figure \ref{figura1} and Figure \ref{figura2} in which the features of the Keplerian ellipse corresponding to the initial conditions of Table \ref{tavola} are shown in blue for comparison.
\begin{figure*}[ht!]
\centering
\begin{tabular}{ccc}
\epsfig{file=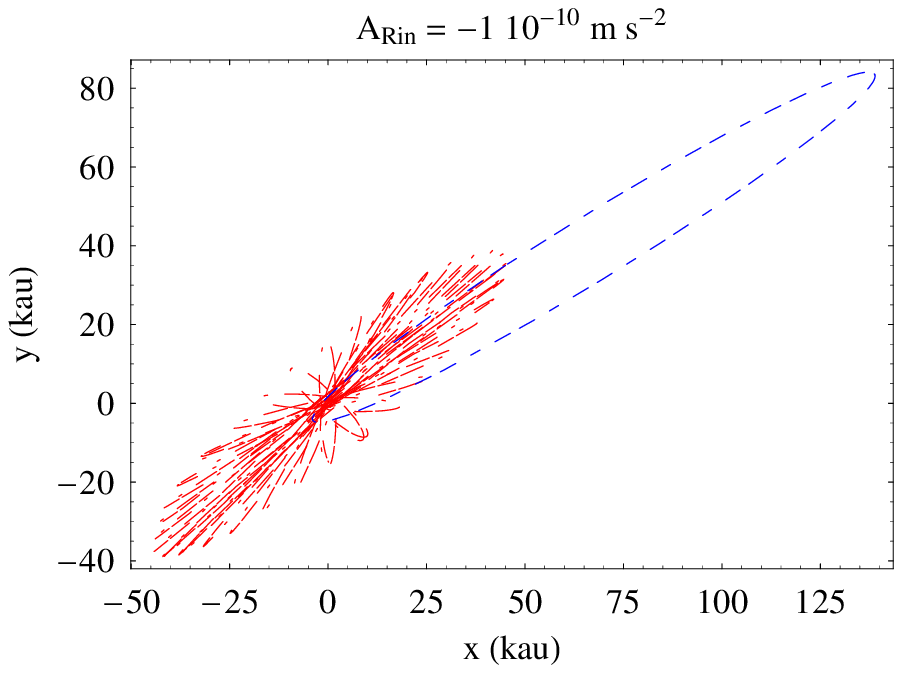,width=0.30\linewidth,clip=} & \epsfig{file=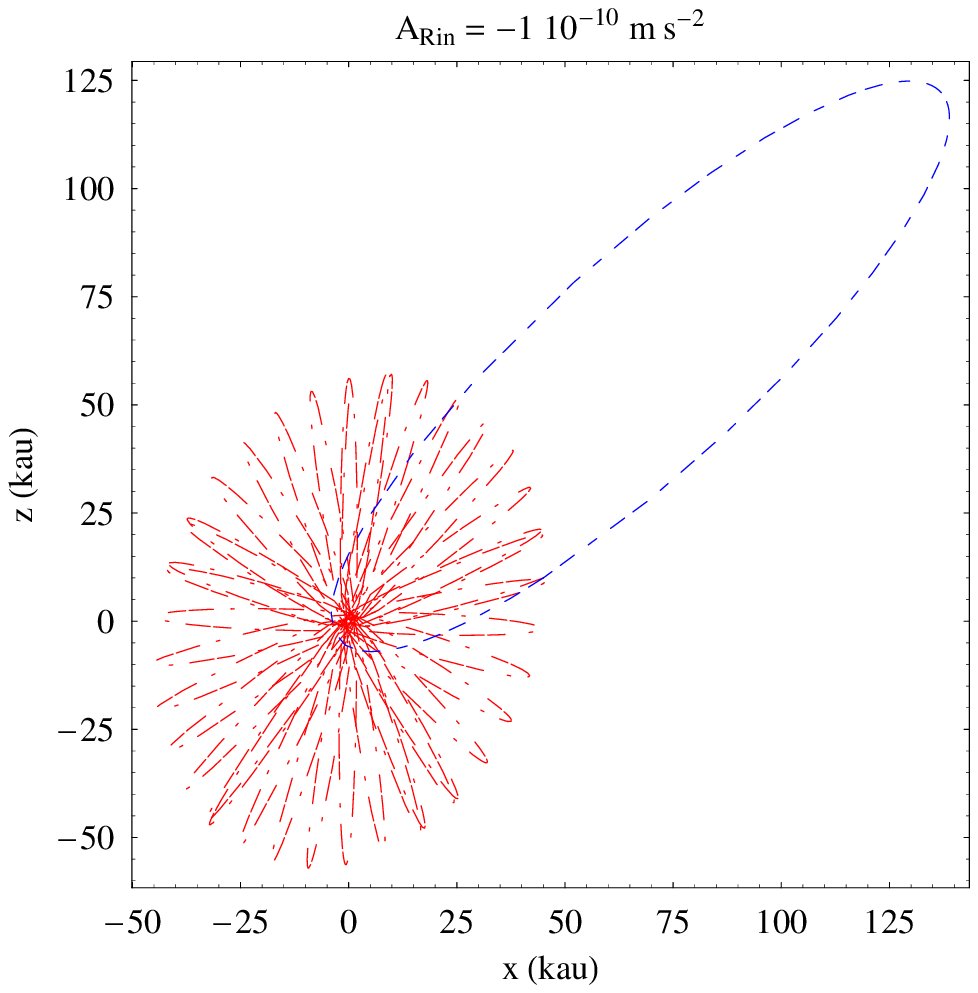,width=0.30\linewidth,clip=}  &
\epsfig{file=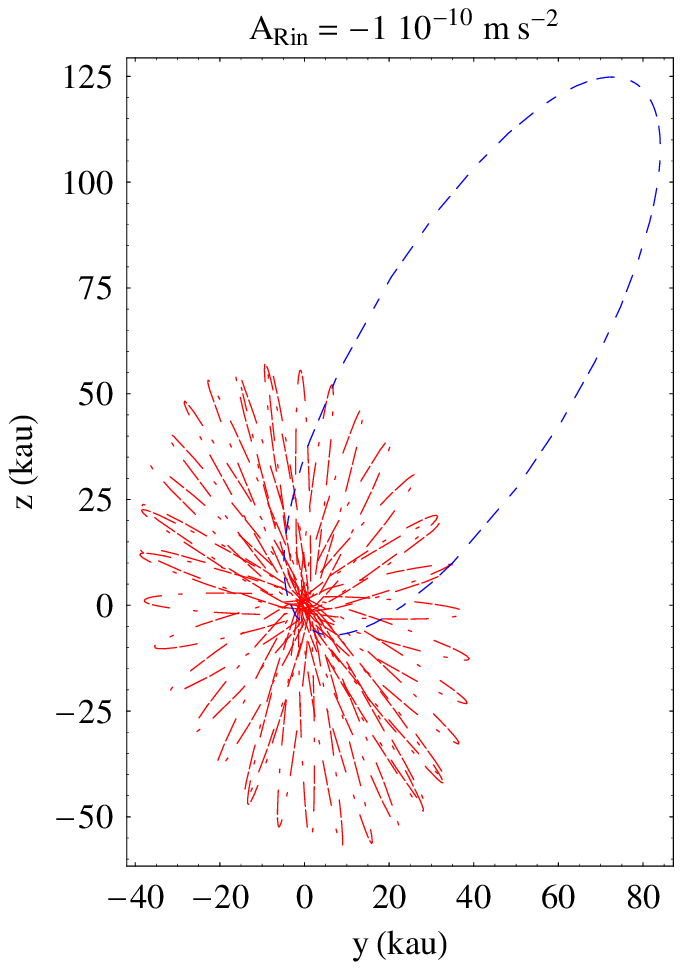,width=0.30\linewidth,clip=} \\
\end{tabular}
\caption{Sections in the coordinates planes of the numerically integrated orbits for the Oort comet of Table \protect{\ref{tavola}} over one Keplerian orbital period $P_{\rm b}$. The dash-dotted red lines  are the projections of the modified orbit  by assuming  that the Rindler acceleration is directed towards the Sun. The dashed blue lines are the projections of the Keplerian ellipse.
}\lb{figura1}
\end{figure*}
\begin{figure*}[ht!]
\centering
\begin{tabular}{ccc}
\epsfig{file=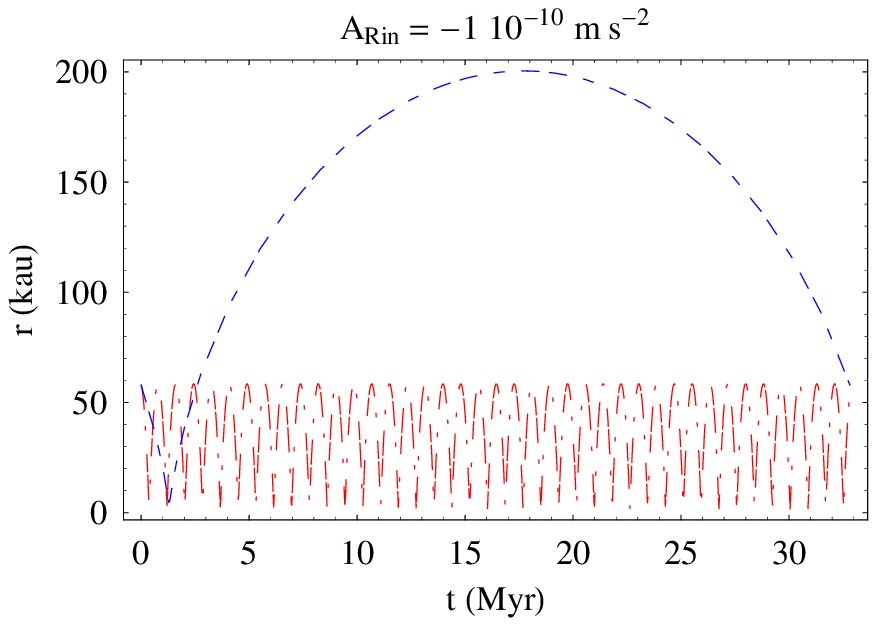,width=0.30\linewidth,clip=} &
\epsfig{file=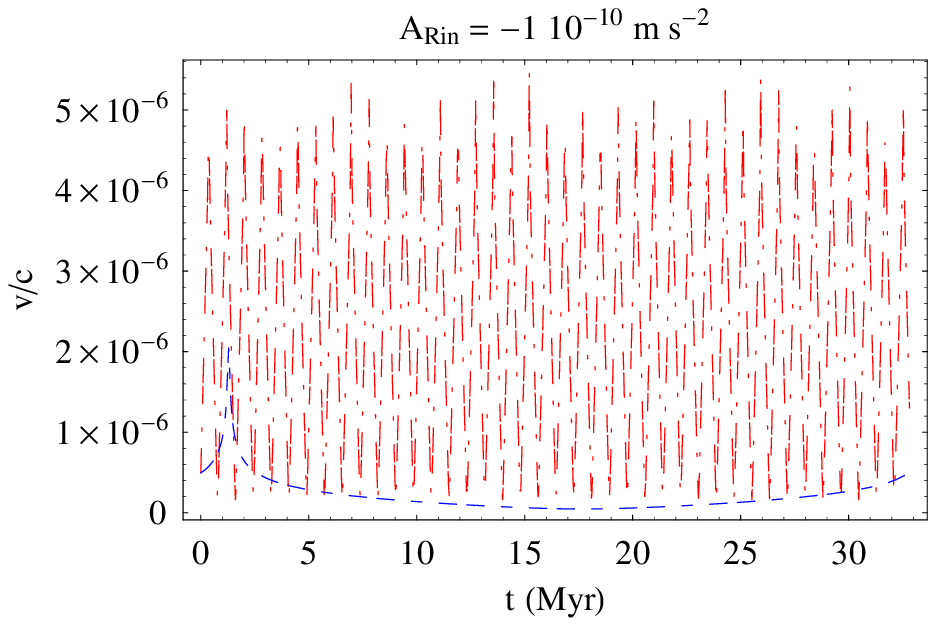,width=0.30\linewidth,clip=} & \epsfig{file=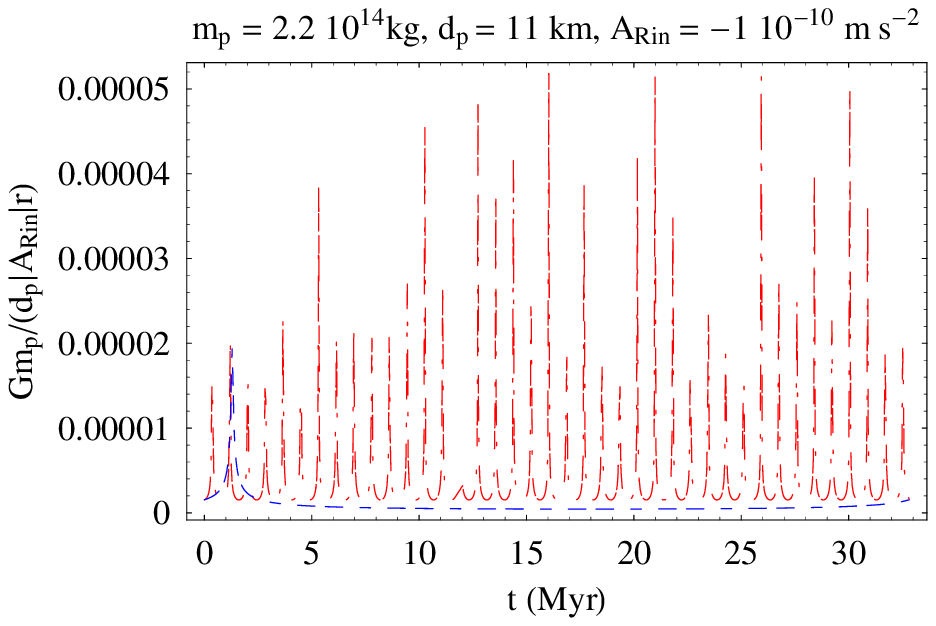,width=0.30\linewidth,clip=}
\end{tabular}
\caption{Heliocentric distances, in kau,  velocities, in units of $c$, and ratio $Gm_{\rm p}/\left(d_{\rm p} \left|A_{\rm Rin}\right| r\right)$ of the numerically integrated orbits for the Oort comet of Table \protect{\ref{tavola}} over one Keplerian orbital period $P_{\rm b}$. The dash-dotted red lines  refer to the modified orbit by assuming  that the Rindler acceleration is directed towards the Sun. The dashed blue lines refer to the Keplerian ellipse.
}\lb{figura2}
\end{figure*}
The qualitative differences with the Newtonian case are striking: bound trajectories still occur, but they radically differ from the Newtonian ones. It is important to note that the condition of \rfr{condizione} is always fulfilled, as shown by the right panel of Figure \ref{figura2}. This is a relevant point since, in principle, it may happen that the comet enters a region in which its gravitational self-energy becomes larger than its putative Rindler energy, thus destroying the validity of the approach followed which was implemented by keeping $A_{\rm Rin}\neq 0$ throughout the whole integration. As anticipated in Section \ref{intro}, the non-relativistic treatment is  justified by the smallness of the comet's speed $v$, displayed in the middle panel of Figure \ref{figura2}. Globally, the heliocentric distance $r$ is strongly reduced, especially in the regions which in the Newtonian case correspond to the aphelion: the minimum distance suffers a relatively smaller reduction with respect to the Newtonian case. Indeed, the left panel of Figure \ref{figura2} tells us that, in the Rindler case, $r$ oscillates with high frequency between about 2 kau and 50 kau.
\subsection{An outward Rindler acceleration}\lb{sec.2.2}
If the Rindler-type acceleration is radially directed from the Sun to the comet, no bounds orbit may exist, as depicted by Figure \ref{figura3} which refers to the same initial conditions of Table \ref{tavola}.
\begin{figure*}[ht!]
\centering
\begin{tabular}{cc}
\epsfig{file=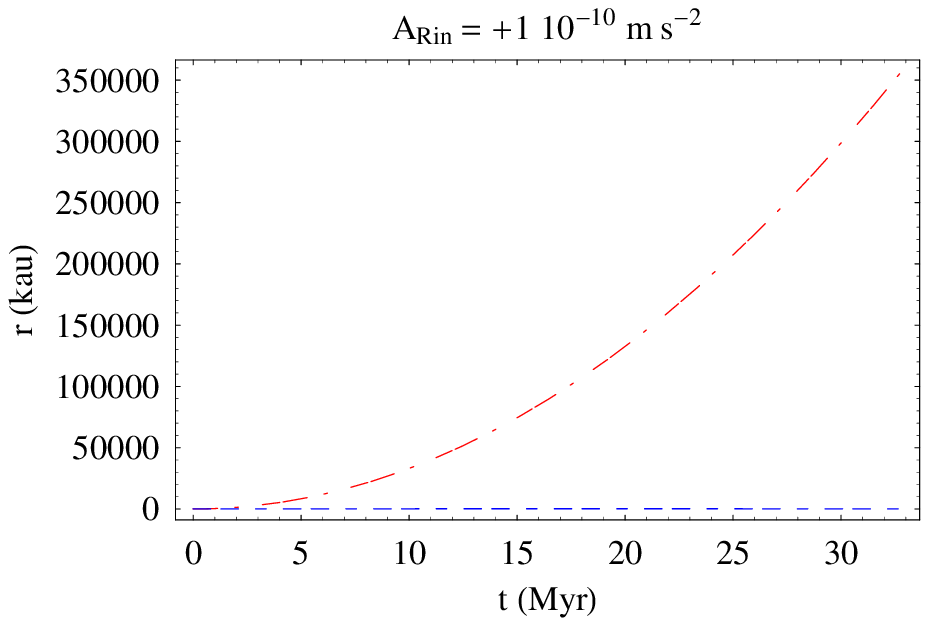,width=0.30\linewidth,clip=} &
\epsfig{file=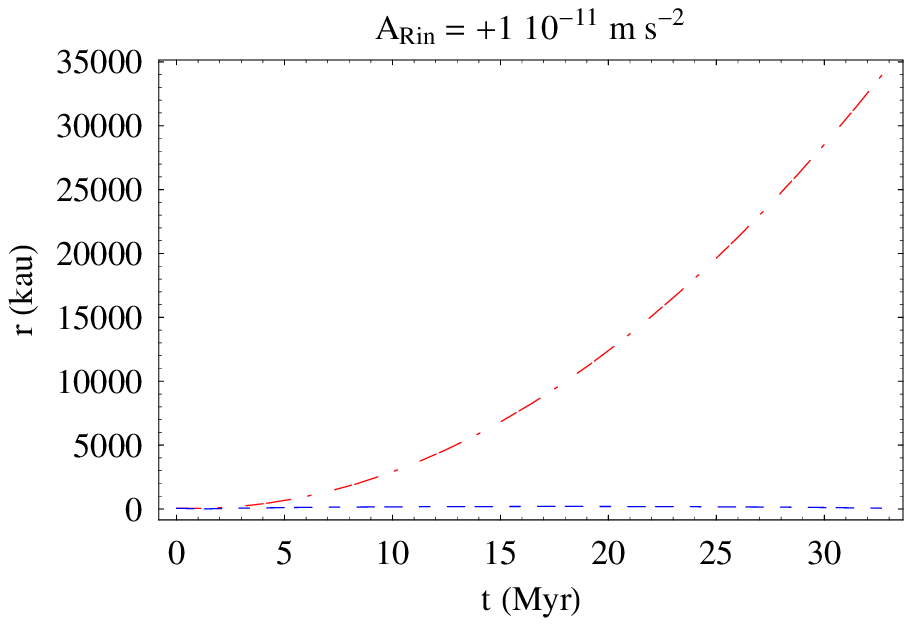,width=0.30\linewidth,clip=} \\
\epsfig{file=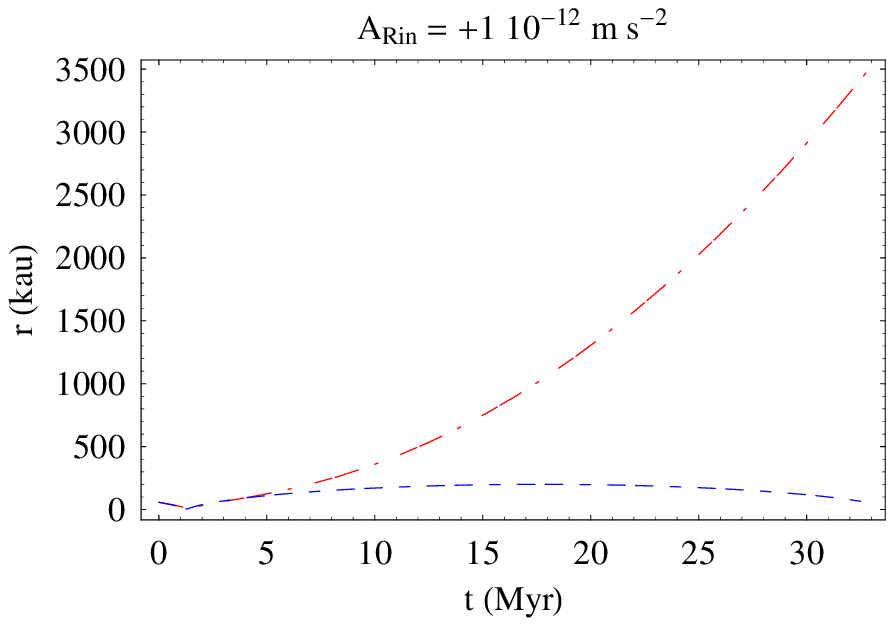,width=0.30\linewidth,clip=} &
\epsfig{file=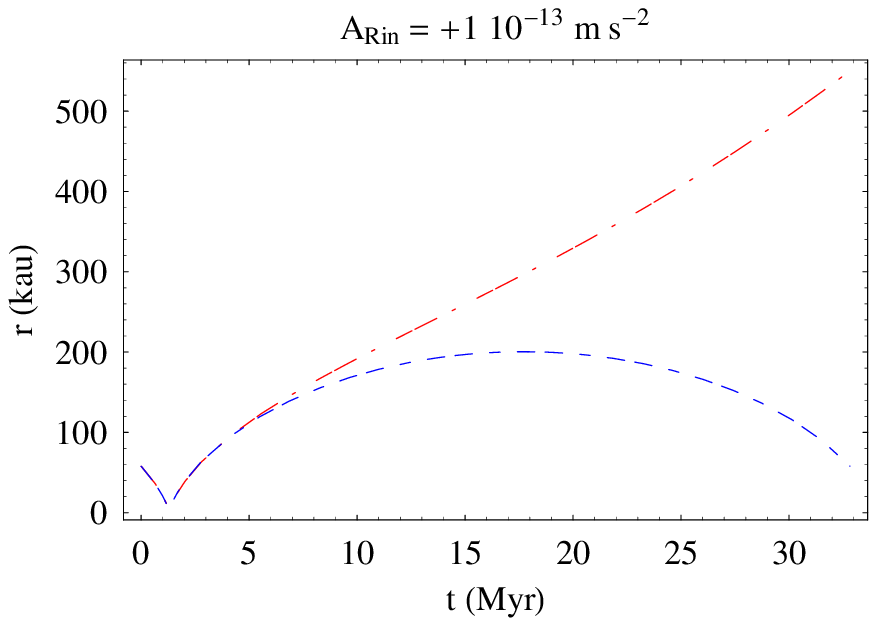,width=0.30\linewidth,clip=}
\end{tabular}
\caption{Numerically integrated heliocentric distances $r$, in kau, of the modified trajectory (dash-dotted red lines) of an Oort comet with the initial conditions of Table \protect{\ref{tavola}} for decreasing positive values of $A_{\rm Rin}$.The duration of the integration is one Keplerian orbital period $P_{\rm b}$.
}\lb{figura3}
\end{figure*}
We adopted decreasing values of $A_{\rm Rin}$ with respect to $\left|A_{\rm Rin}\right|=1\times 10^{-10}$ m s$^{-2}$, but the result is substantially the same.
\section{Consequences of a Rindler-type acceleration on a perturbed Oort cloud}\lb{sec.3}
It seems plausible to expect that the Rindler trajectories are less sensitive than the Newtonian one to those disturbances which affect the Oort cloud yielding those phenomena which are the basis of our confidence in the existence of the cloud with the features usually attributed to it. If so, the existence of the Rindler acceleration would drastically alter the dynamical history of the Oort cloud.
\subsection{The Galactic tide}\lb{sec.3.1}
The effect of the Galactic tide on the orbit of an Oort comet \citep{Heis86} can be obtained
from the following expression of the Galactic tidal acceleration
{
\citep{Fouch05}
\eqi
\begin{array}{lll}
A^{\rm (tid)}_x &=&-\mathcal{G}_1 x^{'}\cos\left(\Omega_0 t\right) + \mathcal{G}_2 y^{'}\sin\left(\Omega_0 t\right), \\ \\
A^{\rm (tid)}_y &=&-\mathcal{G}_1 x^{'}\sin\left(\Omega_0 t\right) - \mathcal{G}_2 y^{'}\cos\left(\Omega_0 t\right), \\ \\
A^{\rm (tid)}_z &=&-\mathcal{G}_3 z. \\ \\
\end{array}\lb{morbi}
\eqf
The various quantities entering  \rfr{morbi} are
\eqi
\begin{array}{lll}
x^{'} & \doteq & x\cos\left(\Omega_0 t\right) + y \sin\left(\Omega_0 t\right), \\ \\
y^{'} & \doteq & -x\sin\left(\Omega_0 t\right) + y \cos\left(\Omega_0 t\right). \\ \\
\end{array}\lb{rotating}
\eqf
and
\eqi
\begin{array}{lll}
\mathcal{G}_1 &\doteq &-\left({\rm A}-{\rm B}\right)\left(3{\rm A} + {\rm B}\right) , \\ \\
\mathcal{G}_2 &\doteq & \left({\rm A}-{\rm B}\right)^2, \\ \\
\mathcal{G}_3 &\doteq & 4\pi G\rho_0-2\left({\rm B}^2-{\rm A}^2\right), \\ \\
\end{array}\lb{oortcon}
\eqf
where A and B are the \citet{Oort27} constants,} $\rho_0$ is the mass density in the solar neighborhood, and $\Omega_0$ is the frequency of solar revolution\footnote{{Since the motion of the Sun about the Galactic Center is
clockwise in the frame adopted \citep{Fouch05}, $\Omega_0$ is negative in it, i.e., it is directed along the $-z$ axis.}}  around the Galaxy.
Both visible stars and Dark Matter concur to the local mass density in the solar neighborhood.
Concerning the baryonic stellar component, the  local mass
density of main-sequence stars is \citep{Reid02} \eqi\rho_{\rm MS}\sim 0.031 M_{\odot}\ {\rm pc}^{-3}=2.09\times 10^{-24}\ {\rm g\ cm^{-3}},\eqf while, according to the most recent results, the Dark Matter local density in the Galactic solar neighborhood is \citep{deBoer011}
\eqi\rho_{\rm DM}=1.3\ {\rm Gev}\ {\rm cm^{-3}}=2.31\times 10^{-24}\ {\rm g\ cm^{-3}}.\eqf
Thus, we will assume
\eqi\rho_0 \sim 4\times 10^{-24}\ {\rm g\ cm^{-3}}=1.5\times 10^{22}\ {\rm kg\ kau^{-3}}.\lb{densita}\eqf

The frequency of the Galactic revolution of the Sun   $\Omega_0$ can be evaluated from its circular rotation speed \citep{Reid09} $\Theta_0=254$ km s$^{-1}$ and its distance  to the Galactic center \citep{Reid09} $R_0=8.4$ kpc as
\eqi\left|\Omega_0\right|=\rp{\Theta_0}{R}=0.0309\ {\rm Myr^{-1}}.\eqf Thus, its period of revolution is
\eqi T_0=203.176\ {\rm Myr}.\eqf

{Concerning the Oort constants A and B entering \rfr{oortcon},
%
%
%
their most recent values, accurate to  $5\%$, are \citep{Feast97}
\eqi
\begin{array}{lll}
{\rm A} & = & 14.82\ {\rm km\ s^{-1}\ kpc^{-1}}=0.0151\ {\rm Myr^{-1}}, \\ \\
{\rm B} & = & -12.37\ {\rm km\ s^{-1}\ kpc^{-1}}=-0.0126\ {\rm Myr^{-1}}. \\ \\
\end{array}\lb{oortcos}
\eqf
}

{
Thus, \rfr{densita} and \rfr{oortcos}, inserted in \rfr{oortcon}, yield\footnote{{Cfr. with the values by \citet{Levi01}.}}
\eqi
\begin{array}{lll}
\mathcal{G}_1 & = & -9.1261\times 10^{-4}\ {\rm Myr^{-2}}, \\ \\
\mathcal{G}_2 & = & 7.7325\times 10^{-4}\ {\rm Myr^{-2}}, \\ \\
\mathcal{G}_3 & = & 3.9626\times 10^{-3}\ {\rm Myr^{-2}}. \\ \\
\end{array}\lb{tidalco}
\eqf
}

In Figure \ref{figura4} we show the numerically integrated Rindlerian and Newtonian trajectories, including \rfr{morbi}, over a full revolution of the Sun around the Galaxy. We used the same initial conditions of Table \ref{tavola} for the Oort comet.
\begin{figure*}[ht!]
\centering
\begin{tabular}{cc}
\epsfig{file=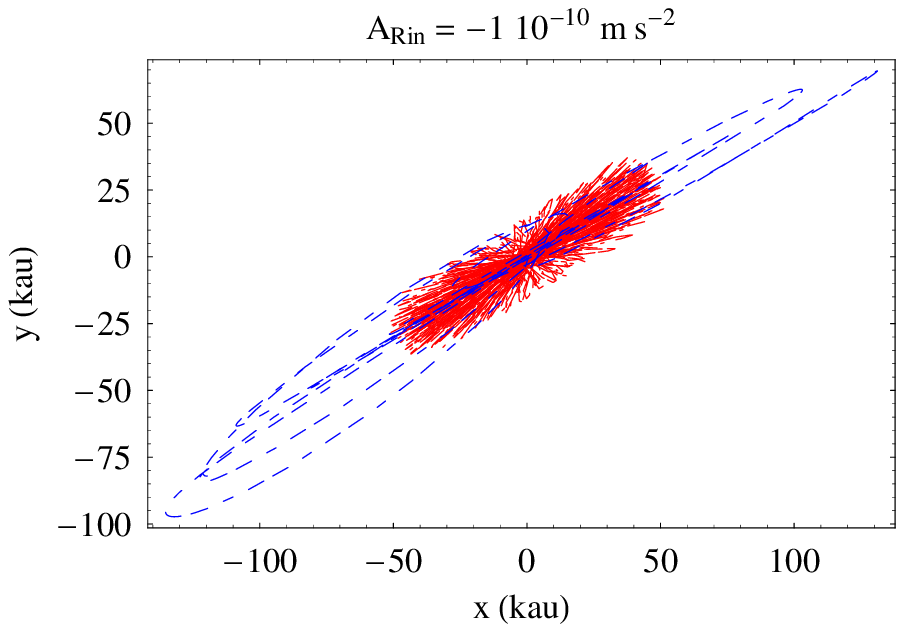,width=0.30\linewidth,clip=} &
\epsfig{file=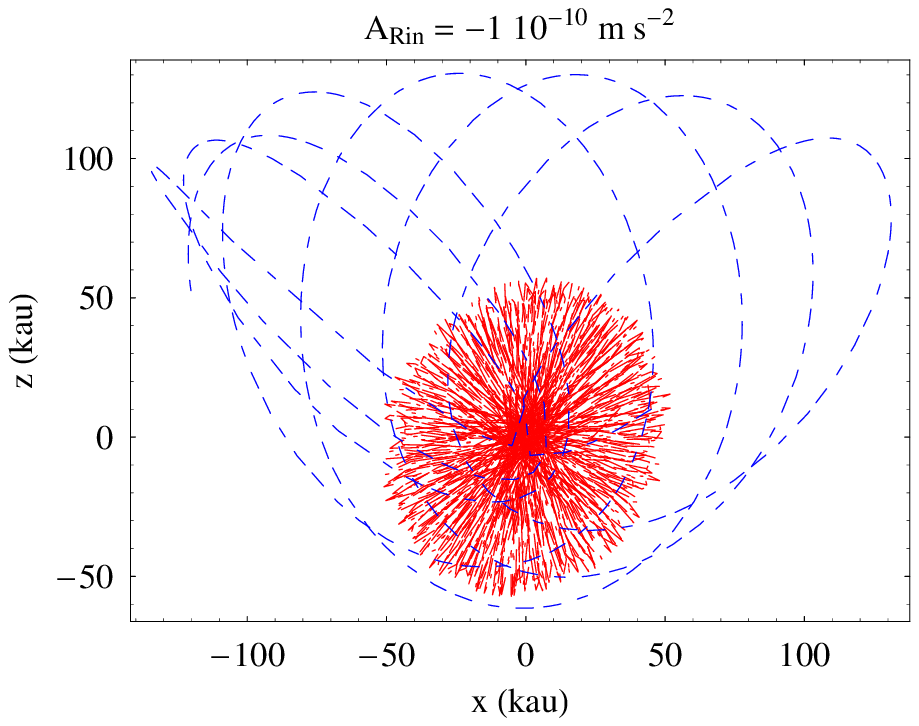,width=0.30\linewidth,clip=} \\
\epsfig{file=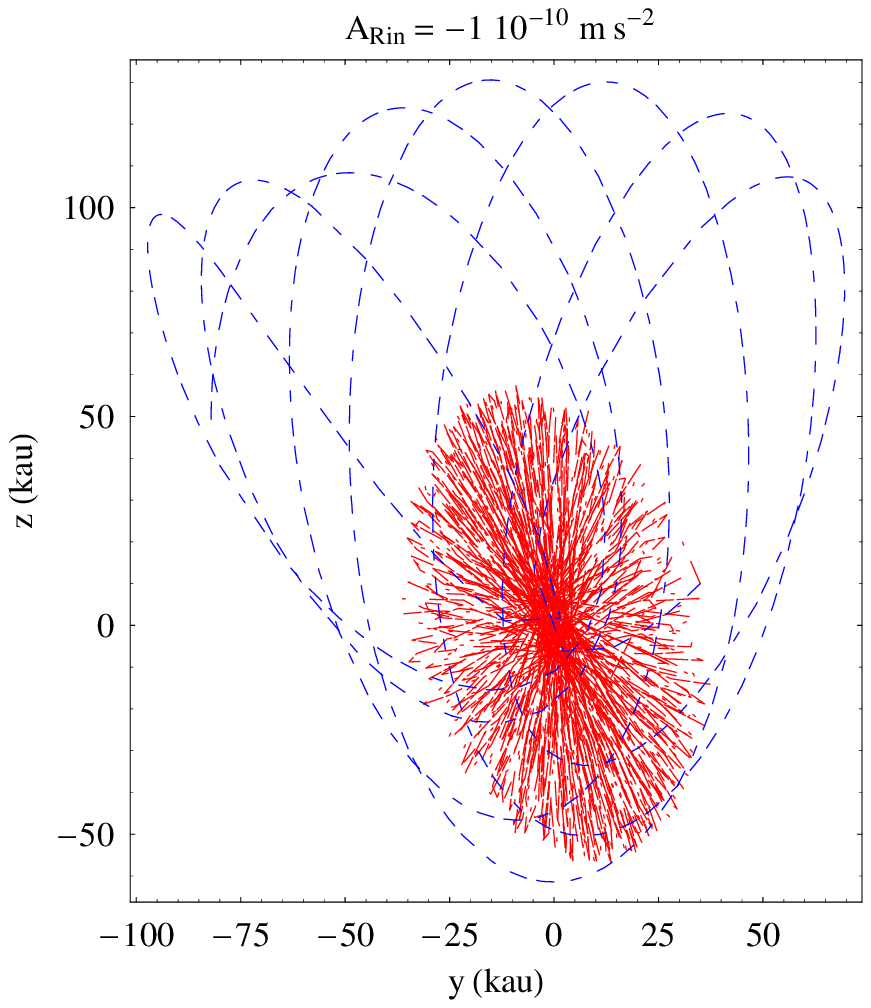,width=0.30\linewidth,clip=} &
\epsfig{file=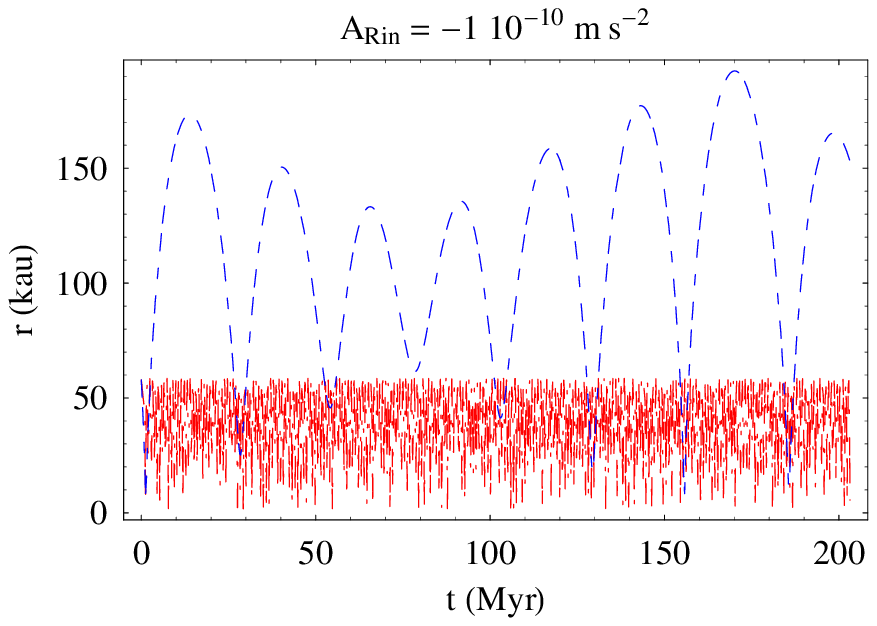,width=0.30\linewidth,clip=}
\end{tabular}
\caption{Coordinate sections of the numerically integrated trajectories and heliocentric distances $r$ of an Oort comet affected by the Galactic tide of \rfr{morbi} with (dash-dotted red lines) and without (dashed blue lines) the Rindler-type acceleration of \rfr{magni}. The initial conditions of the Oort comet are those of Table \protect{\ref{tavola}}. The duration of the integration is one full Galactic revolution ($T_0=203.176$ Myr).
}\lb{figura4}
\end{figure*}
As expected, the differences between the Newtonian and the modified scenarios are remarkable: a comparison with Figure \ref{figura1} and Figure \ref{figura2} shows that the Rindler trajectory is left substantially unaffected by the Galactic tide, contrary to the Newtonian one. It can be shown that also for an outward Rindler acceleration the Galactic tide does not have influence in the sense that the comet does not remain bound, as in Figure \ref{figura3}.
\subsection{A close encounter with a passing star}\lb{sec.3.2}
Here we consider the perturbing action of a passing star s with $m_{\rm s}=M_{\odot}$ over an unperturbed Keplerian orbital period $P_{\rm b}$ of the Oort comet.
The initial conditions of s are listed in Table \ref{tavola2}.
\begin{table*}[ht!]
\caption{Initial conditions, in kau and ${\rm kau\ Myr^{-1}=4.74\ m\ s^{-1}}$, of a perturbing passing star s adopted for the numerical integration of the modified equations of motion of an Oort comet acted upon by the Rindler-type acceleration of \rfr{grum}. They correspond to a heliocentric distance  $r_0^{\rm (s)}=3.2$  pc and a speed  $v_0^{\rm (s)}=0.61$ km s$^{-1}$.
}\label{tavola2}
\centering
\bigskip
\begin{tabular}{llllll}
\hline\noalign{\smallskip}
$x^{\rm (s)}_0$ (kau) & $y^{\rm (s)}_0$ (kau) & $z^{\rm (s)}_0$ (kau) &  $\dot x^{\rm (s)}_0$ (kau Myr$^{-1}$) & $\dot y^{\rm (s)}_0$ (kau Myr$^{-1}$) & $\dot z^{\rm (s)}_0$ (kau Myr$^{-1}$) \\
\noalign{\smallskip}\hline\noalign{\smallskip}
$270$ & $370$ & $470$ & $-42.18$ &  $-63.28$ & $-105.47$\\
\noalign{\smallskip}\hline\noalign{\smallskip}
\end{tabular}
\end{table*}
In the numerical integration of the equations of motion we assume that s moves uniformly with $v^{\rm (s)}(t)=v_0^{\rm (s)}$.
The result is depicted in Figure \ref{figura5}.
\begin{figure*}[ht!]
\centering
\begin{tabular}{cc}
\epsfig{file=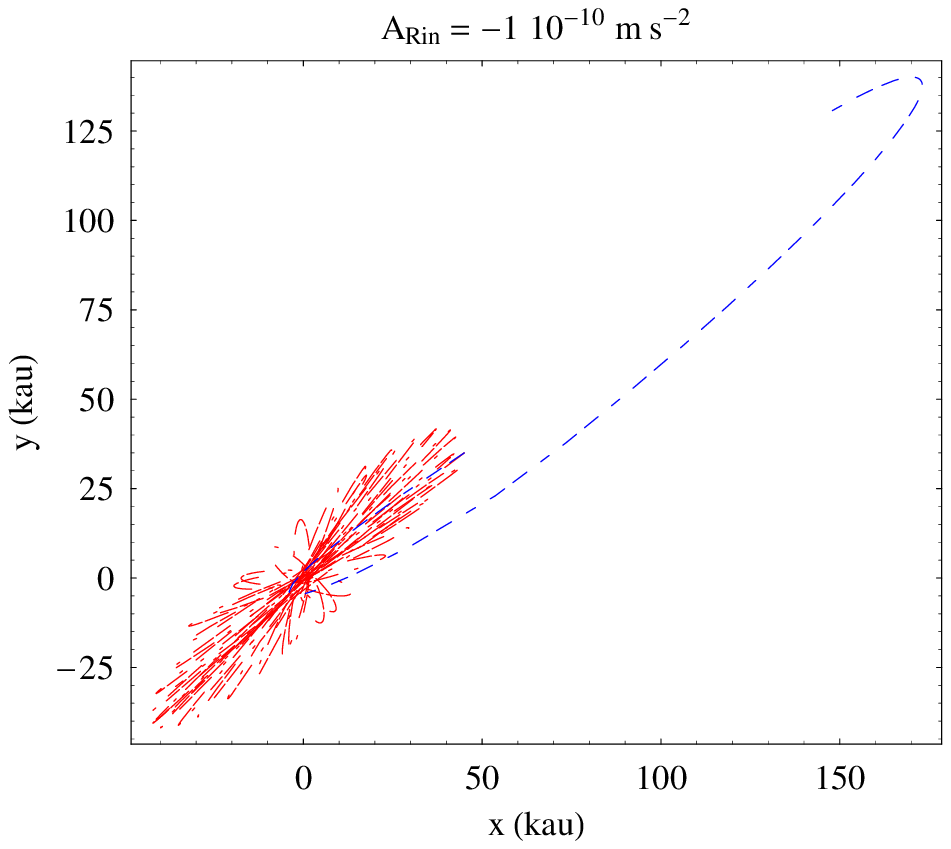,width=0.30\linewidth,clip=} &
\epsfig{file=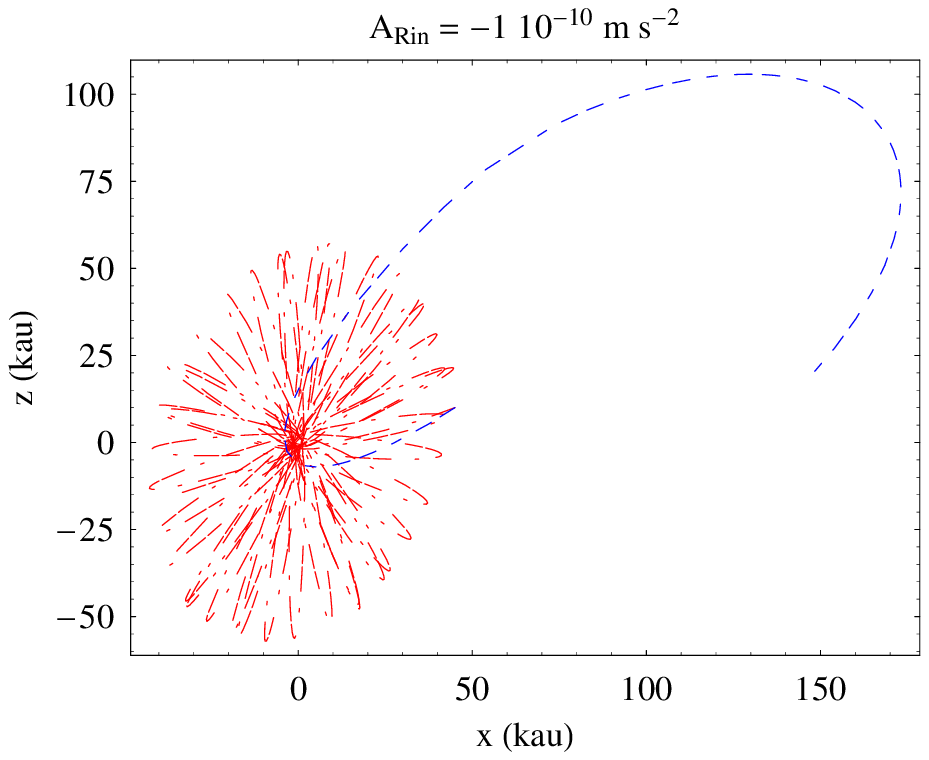,width=0.30\linewidth,clip=} \\
\epsfig{file=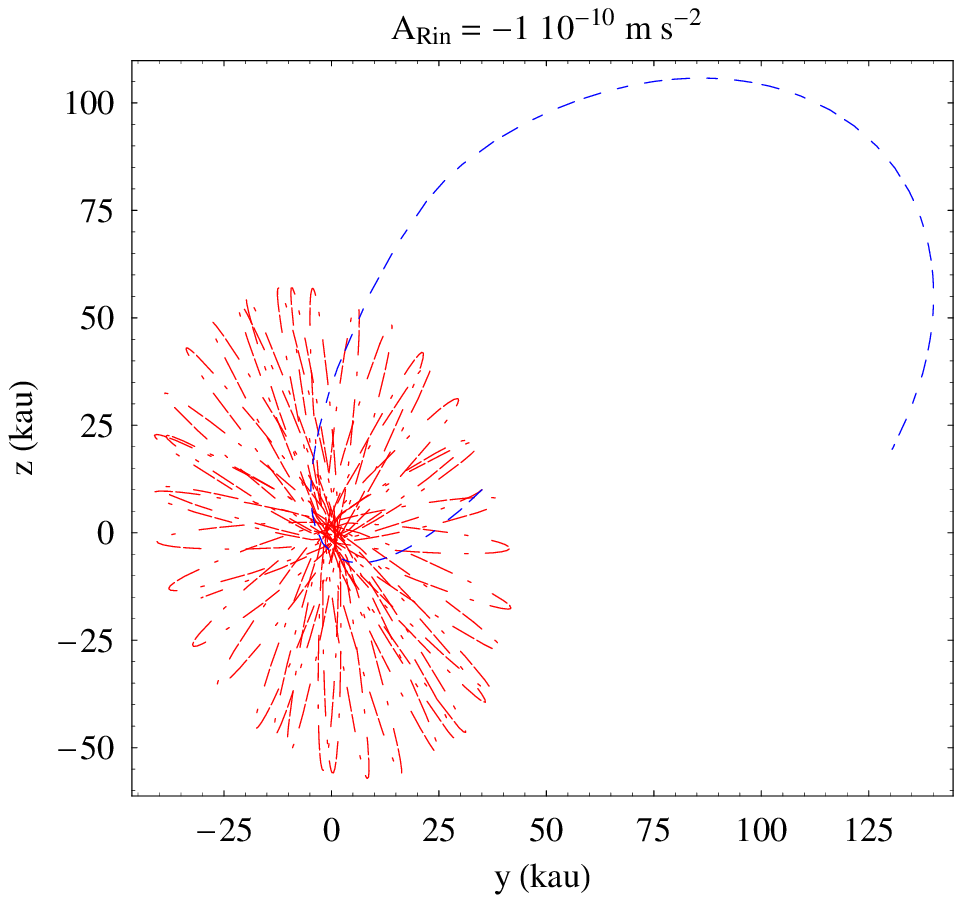,width=0.30\linewidth,clip=} &
\epsfig{file=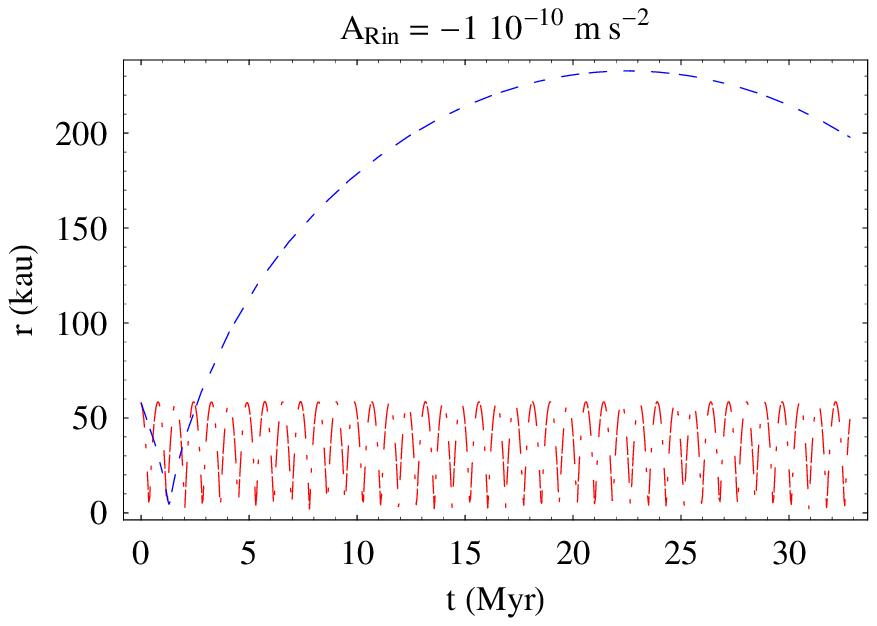,width=0.30\linewidth,clip=}
\end{tabular}
\caption{Coordinate sections of the numerically integrated trajectories and heliocentric distances $r$ of an Oort comet affected by the passing star of Table \protect{\ref{tavola2}}  with (dash-dotted red lines) and without (dashed blue lines) the Rindler-type acceleration of \rfr{magni}. The initial conditions of the comet are those of Table \protect{\ref{tavola}}. The duration of the integration is one (unperturbed) Keplerian period $P_{\rm b}$.
}\lb{figura5}
\end{figure*}
It can be noticed that, while the Keplerian ellipse is, actually, distorted\footnote{In this particular case, the comet is stripped away by the stellar passage: indeed, this is one of the causes of the gradual depletion of the Oort cloud \citep{Remy85}.} by the Newtonian interaction with s, the trajectory computed with the Rindler acceleration remains substantially unaffected. Also in this case, it can be shown that no bound orbits can exist if $\bds A_{\rm Rin}$ is directed outward.
\section{Discussion and conclusions}\lb{sec.4}
The analysis performed in this paper should be considered just as a preliminary one. It aims at exploring semi-qualitatively how the presence of an additional radial Rindler-like acceleration may affect the orbital motion of  bodies whose self-energy is smaller than their putative Rindler energy like the objects moving in the Oort cloud. We only choose a specific set of initial conditions which correspond, in the Newtonian scenario, to a typical orbital configuration for an Oort comet moving on a highly eccentric and inclined ellipse which almost extends throughout the entire expected extension of the Oort cloud. More refined numerical analyses may explore  larger ensembles of initial conditions with a statistical approach. It would also be interesting to repeat Monte Carlo simulations of the dynamical evolution of the  Oort cloud over the age of the solar system like, e.g., those by \citet{Rick08} and \citet{Masi09} by explicitly including a Rindler-type extra-acceleration, and inspect how several key features of the cloud change with respect to the usual Newtonian scenario.

In our numerical analysis we found that the standard Newtonian picture is notably altered by an additional radial Rindler acceleration as large as $10^{-10}$ m s$^{-2}$ which, in the case of an Oort comet, would not be smaller than the Newtonian one. Bound orbits are not possible if $\bds A_{\rm Rin}$ is directed outward, even for $A_{\rm Rin}=10^{-13}$ m s$^{-2}$. Conversely, if the Rindler acceleration is directed from the comet to the Sun the resulting trajectory is limited in space, but it is completely different from a Keplerian ellipse. In particular, the heliocentric distance $r$ is greatly reduced, and it experiences high-frequency variations during a Keplerian orbital period $P_{\rm b}$. Moreover, the spatial pattern of the modified trajectory is quite isotropic over $P_{\rm b}$. As a consequence, the Rindlerian trajectory is much less sensitive than the Newtonian one to disturbing effects like the Galactic tide and nearby passing stars.

Although, as pointed out above, more extended numerical investigations should be implemented by varying the initial conditions in a Monte Carlo fashion, it is difficult to believe that the main features of the Oort cloud, which lead to a general consensus about its existence, may be preserved by the existence of a Rindler-type extra-acceleration of the order of $10^{-10}$ m s$^{-2}$.
Our results are not necessarily limited to the specific model by \citet{Gru010}: indeed, they are quite general and are valid for any hypothetical constant and uniform radial acceleration. In the case of the Pioneer anomaly,  the magnitude of the anomalous acceleration may be as large as $\left|A_{\rm Pio}\right|\leq 1\times 10^{-9}$ m s$^{-2}$ and all the previous results remain qualitatively valid, being the anomalous behavior of an Oort comet even more remarkable from a quantitative point of view. Finally, we notice that using the Oort cloud may, in principle, be useful also for other long-range modified models of gravity.
{
\section*{Acknowledgments}
I thank an anonymous referee for a pertinent critical remark.}


\end{document}